# Phosphatidylserine transport in cell life and death

Alenka Čopič[1,*], Thibaud Dieudonné[2], Guillaume Lenoir[2]

[1]Centre de Recherche en Biologie Cellulaire de Montpellier (CRBM), Université de Montpellier, CNRS, 34293 Montpellier CEDEX 05, France

[2]Université Paris-Saclay, CEA, CNRS, Institute for Integrative Biology of the Cell, 91198 Gif-sur-Yvette, France

*Correspondence: alenka.copic@crbm.cnrs.fr

## Abstract

Phosphatidylserine (PS) is a negatively-charged glycerophospholipid found mainly in the plasma membrane (PM) and in the late secretory/endocytic compartments, where it regulates cellular activity and can mediate apoptosis. Export of PS from the endoplasmic reticulum, its site of synthesis, to other compartments, and its transbilayer asymmetry must therefore be precisely regulated. We review recent findings on non-vesicular transport of PS by lipid transfer proteins (LTPs) at membrane contact sites, on PS flip-flop between membrane leaflets by flippases and scramblases, and on PS nano-clustering at the PM. We also discuss emerging data on cooperation between scramblases and LTPs, how perturbation of PS distribution can lead to disease, and the specific role of PS in viral infection.

## Introduction

Phosphatidylserine (PS) is a negatively-charged phospholipid with a highly uneven distribution in cellular membranes, which is important for cellular function [1-3]. It is also used as a signal for engulfment of apoptotic cells by macrophages, and is exploited by pathogens for cell entry and propagation [4]. First, although PS is synthesized at the endoplasmic reticulum (ER), it is enriched in the plasma membrane (PM) and in late secretory/endocytic compartments. Second, an even steeper gradient exists between the two leaflets of the PM, as well as the *trans*-Golgi network (TGN) and endosomes, where PS is confined to the cytosolic side. Third, at the PM and possibly also in other compartments, PS has been shown to laterally segregate into nanodomains. Such uneven distribution necessitates mechanisms of selective PS transport and/or retention and depletion from specific membrane compartments. Here, we highlight some recent advances in our understanding of the function of LTPs, which can mediate PS transport between compartments, and of lipid flippases and scramblases, which mediate its flip-flop between membrane leaflets. We specifically focus on the selectivity of these proteins for PS, and how the selectivity can be increased by other processes.

**PS transport between cellular compartments**

Like most other phospholipids, PS is synthesized at the ER, where its synthesis is under feedback inhibition [3]. LTPs, functioning within membrane contact sites, can extract PS from the cytosolic leaflet of the ER to transport it to other compartments [5,6]. Alternatively, PS can exit the ER following bulk membrane flow via COPII-coated membrane carriers.

In contrast to vesicular transport, it is well established that some LTPs can transport PS in a highly selective manner. This was first shown for two homologous yeast proteins from the Osh/ORP family, Osh6 and Osh7, which localize to the ER-PM contact sites, and later for their orthologues ORP5 and ORP8 in mammalian cells [7,8]. These proteins, which are related via their lipid-binding ORD domain, exchange PS with phosphatidylinositol-4-phosphate (PI4P). PI4P is continuously generated at the PM and dephosphorylated back to phosphatidylinositol at the ER, thus fueling the transport of PS against its concentration gradient [8,9]. *In vitro*, Osh6 and ORP8 show preference for unsaturated PS species [10], suggesting that they may contribute to divergent acyl chain composition of different cellular compartments.

Three other members of the ORP family, ORP9, ORP10 and ORP11, which are most closely related to ORP5/8, localized to the ER-TGN contact sites and influenced PS levels at the TGN [11]. Two recent studies further show that they mediate PS-PI4P exchange between the ER and the endocytic compartments to support downstream processes, either fission of recycling endosomes [12], or lysosomal repair [**13]. The lysosomal repair pathway is triggered by accumulation of PI4P in response to lysosome damage, followed by recruitment ORP9/10/11 and an increase in PS on the lysosomal membrane, which in turn leads to the recruitment of another LTP, ATG2A (Figure 1). ATG2A belongs to the Vps13-related family of large channel-like LTPs [14], which can transport phospholipids in a very efficient but non-selective manner; the activity of ATG2A could therefore enable rapid sealing of the damaged lysosomes [**13]. Cholesterol, supplied by the ORP family members OSBP and ORP1L, is also required for lysosomal repair [13,15]; the details of this process remain to be explored.

In yeast, Osh6/7 are the only known LTPs with a clearly demonstrated preference and role in PS transport; however, in cells lacking Osh6/7, enrichment of PS at the PM is still observed at steady state, suggesting additional mechanisms. Tricalbins, members of the extended synaptotagmin LTP family that localize to the ER-PM contacts, maintain PS levels at the PM under heat stress conditions [16]; the underlying mechanisms remain to be explored. Enrichment of PS in a target membrane can also be achieved by coupling a non-specific lipid transport pathway with specific retention of PS in the target membrane, for example by virtue of its association with other lipids or proteins, as discussed in the final chapter. At the donor site, lipid export can be promoted by lipid synthesis [17], but direct evidence for a positive feedback between the two processes is currently lacking. Opposite to this, the existence of a PS sink in the target compartment, *i.e.*, a mechanism to deplete PS upon delivery, can create a downward gradient to promote PS flow. This may be important for PS transport to mitochondria, where PS is converted to phosphatidylethanolamine (PE) via decarboxylation, following transfer by a specific intra-mitochondrial LTP complex to the inner mitochondrial membrane [5,18,19]. In yeast, a second PS-decarboxylase, Psd2, localizes to endosomes and functions downstream of Osh6/7 [20]; it may therefore represent a mechanism to regulate PS levels at the PM via endocytosis.

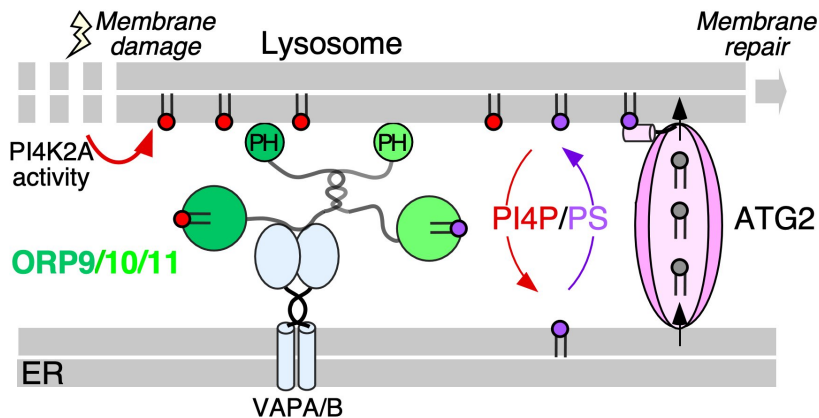

***Figure 1. Partnership of LTPs during lysosomal membrane repair.*** *Lysosome membrane damage induces the production of PI4P by the PI4-kinase PI4K2A. PI4P accumulation promotes the formation of ER-lysosome membrane contacts through recruitment of ORP9, which binds to VAP proteins in the ER membrane. ORP10 and ORP11 are also recruited to this contact site likely via formation of hetero-dimers with OPR9, because they don't display a strong affinity for VAP proteins. These ORPs mediate the exchange of PI4P with PS and thereby accumulation of PS at the lysosomal membrane, which in turn promotes the binding of ATG2 via an amphipathic helix. High phospholipid transfer activity of ATG2 enables repair of the lysosomal membrane, likely in collaboration with LTPs that transport cholesterol (not shown) [11-13,15].*

By what route PS reaches the mitochondria has been the subject of intense investigations; non-selective glycerophospholipid transporters, such as Vps13 and the ERMES complex, play an important role [21]. PS-specific LTPs have also been observed at the ER-mitochondria contact sites; in particular, ORP5/8 are enriched in these contacts and affect transport of PS in cells [22]. Mitochondria lack PI4P, therefore there can be no counter-exchange of PI4P in this case, consistent with the down-hill flow of PS. Interestingly, the Drin group has observed that the acyl-chain specificity of the ORP8-ORD- or Osh6-mediated PS transfer *in vitro* changes depending on the presence of PI4P in acceptor membranes, suggesting the possibility of sorting of PS species between the PM and the mitochondria [10]. In yeast cells, PS transport to mitochondria also displays selectivity at the level of acyl chains, with a preference for shorter and more unsaturated PS species [23].

ORP5 and ORP8 have also been detected in contact sites between the ER and the lipid droplets (LDs) that are closely associated with mitochondria, and LDs are affected by ORP5/8 depletion [24]. However, how these phenotypes relate to the PS transport activity of ORP5/8 is not clear. Another recently-identified LTP, mitoguardin-2, also localizes to these tri-partite ER-LD-mitochondria contact sites and was initially shown to have some specificity for PS, but a concomitant study instead suggests a lack of selectivity between different glycerophospholipids [25,26].

**Transbilayer transport of PS**

A very steep PS gradient exists between the two leaflets of the PM, the TGN or the endosomes [27,28], suggesting tight regulation. At the PM, all glycerophospholipids except phosphatidylcholine (PC) are almost exclusively localized to the inner leaflet, whereas the outer leaflet is enriched in PC and sphingolipids. Phospholipid flippases (P4-ATPases in eukaryotic cells) are active transporters that pump lipids towards the cytosolic leaflet of cell membranes. P4-ATPases form heteromeric complexes with members the Cdc50 protein family, which are required for proper localization of the complex (Figure 2). In various instances, such as during blood coagulation or apoptosis, PS must be rapidly flipped to the cell surface, where it serves as a scaffold for blood-clotting enzymes or as an 'eat me' signal for clearance by macrophages, a process catalyzed by lipid scramblases [4]. Changes in PS transbilayer distribution are also exploited by various intracellular pathogens (Box 1).

---

**Box 1. Emerging roles of scramblases, flippases and PS distribution in infectious diseases**

Several lines of evidence point to a crucial role of PS transbilayer distribution in host-pathogen interactions. The ATP2B/CDC50.4 lipid flippase complex, which has been shown to transport PS, is required for the lytic cycle of Toxoplasma gondii [68]. In the host cell, the scramblases VMP1 and TMEM41B play a role in the formation of double membrane vesicles (DMV) upon infection by β-coronavirus. This function is dependent on the ability of cells to regulate PS distribution, as depletion of the PS synthase PTDSS1 rescues the DMV defects in VMP1 knock-out cells [69]. Furthermore, analysis of pneumocytes from COVID-19 patients revealed the presence of multinucleated cells called syncytia, and screening for drugs that block SARS-CoV-2 infection yielded molecules targeting TMEM16F. Inhibition of TMEM16F reduced externalization of PS in cells treated with the Ca-ionophore ionomycin and downregulation of TMEM16F blunted syncitia formation in spike protein-expressing cells [70]. These results are consistent with a role of PS in cell fusion events, and are supported by a study showing that compounds inhibiting TMEM16F-mediated PS externalization have antiviral effects [71]. Genetic screens also identified TMEM41B and CDC50A as crucial host factors for infection by human coronaviruses, including SARS-CoV-2 [72]. Furthermore, co-expression of Ebola virus matrix protein VP40 (eVP40) and glycoprotein GP in cells yielded virus-like particles (VLPs) with PS exposed at the surface as a result of incorporation of Xkr8 in VLPs, which increased the uptake of VLPs by target cells in a process known as apoptotic mimicry [73]. Expression of eVP40 was shown to enhance PS clustering at the PM of living cells, which in turn promoted eVP40 membrane binding and oligomerization. Consistently, fendiline, a recently FDA-approved drug that reduces PS levels and clustering at the PM, decreased eVP40 oligomerization and VLP production from infected cells [74].

---

In contrast to early reports suggesting that flippases were PS/PE transporters, only a subset of P4-ATPases actually transports PS, among which mammalian ATP11A, ATP11C, ATP8A1, ATP8A2 and ATP8B1, as well as Drs2 in yeast [29]. Erythrocytes of ATP11C-deficient mice display elevated PS at the cell surface, abnormal shape and reduced lifetime [30]. Mutations in human ATP8A2 that disrupt PS-activated ATPase activity have been reported to cause severe neurological disorders [31]. Interestingly, recent work indicates that elevated PS levels at synapses promote synaptic pruning, i.e., removal of synapses to establish proper connections during brain development, and points to a prominent role of the P4-ATPase chaperone CDC50A in mediating PS exposure and subsequent aberrant synapse removal *in vivo* [32,33]. Similarly, the expression the lipid scramblase Xkr8 has been shown to be dynamically regulated in the developing brain and to contribute to axonal pruning [34]. Intriguingly, the brain-specific G protein-coupled cell adhesion receptor BAI1, originally described to bind PS exposed on apoptotic cells for engulfment by macrophages, is also enriched in the post-synaptic density, where it interacts with ATP11A. The lipid flippase activity of ATP11A reduces the signaling activity of BAI1 in a manner dependent on the extracellular PS-binding motif of BAI1, suggesting a complex mechanism of PS sensing, whether BAI1 is located on microglia/astrocytes or on neurons [*35].

Recent structural investigations have allowed detailed understanding of P4-ATPase-catalyzed PS transport across cell membranes and disclosed the regulatory mechanisms at play [36-39]. ATP8A1, ATP8B1 and the yeast Drs2 are autoinhibited by their C- and N-terminal extensions. For human ATP8B1, phosphorylation of a serine in the C-terminus may control the dissociation of the inhibitory tail [39]. Furthermore, phosphoinositides have been identified as critical activators of Drs2 and ATP8B1 via an unknown molecular mechanism (Figure 2) [36,39,40].

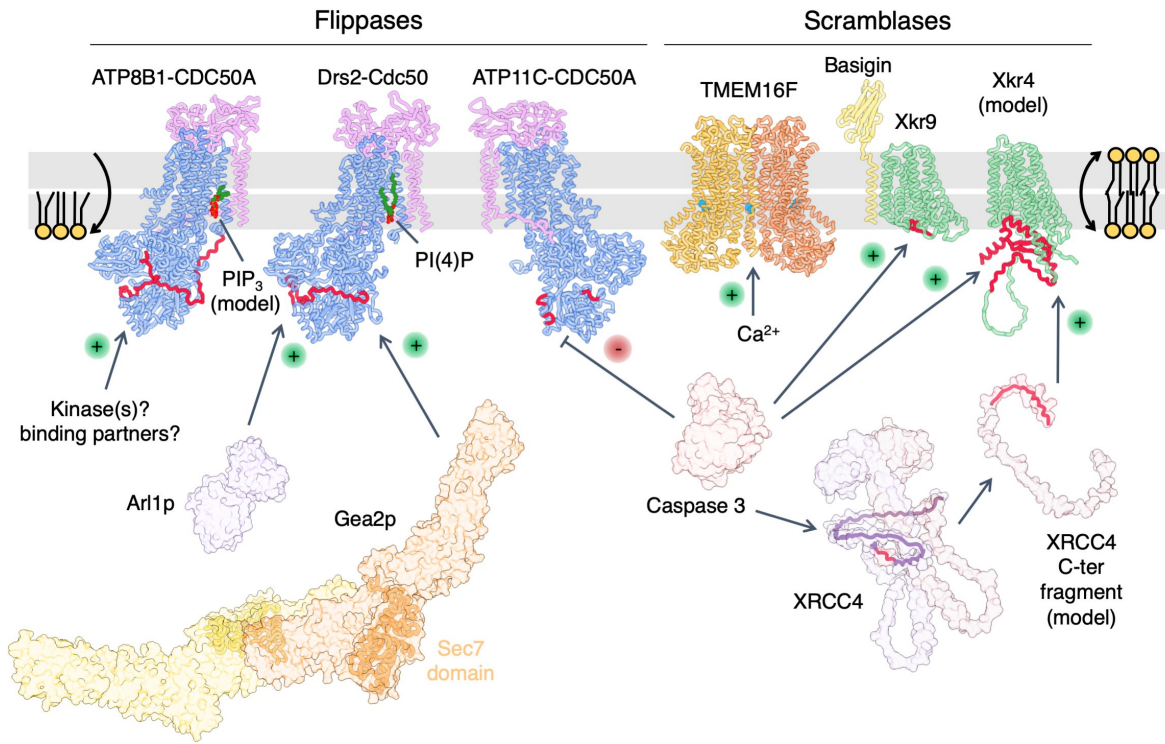

***Figure 2. Regulatory mechanisms underpinning flippase- and scramblase-mediated phospholipid transport.*** *Flippases are exemplified by the human ATP8B1-CD50A (7PY4), yeast Drs2-Cdc50 (6ROJ) and human ATP11C-CDC50A (6LKN) complexes, with the P4-ATPase subunit shown in light blue and Cdc50 proteins in pink. Scramblases are exemplified by human TMEM16F (6QP6), the human Xkr8-Basigin complex (7DCE) and a model of Xkr4 generated with AlphaFold2. Autoinhibition in ATP8B1-CDC50A and Drs2-Cdc50 complexes is achieved through binding of their C- and N-terminal extensions (in red) at the interface between the catalytic ATPase domains, thereby preventing conformational changes required for lipid transport. The mechanism by which autoinhibition is relieved is still unclear, but the N- and C-terminal tails of Drs2 contain binding sites for the small GTP-binding protein Arl1 (1MOZ) and the Sec7 domain (1KU1) of the large Arf-GEF Gea2 (7UTH), respectively. The regulatory phosphoinositides, PI(3,4,5)P$_3$ and PI4P, are displayed in the structures of ATP8B1 and Drs2, respectively. The TMEM16F dimer (yellow/orange), is shown in a Ca$^{2+}$-bound conformation (Ca$^{2+}$ spheres in blue). Ca$^{2+}$ binding stimulates lipid transport by TMEM16F. The Xkr8 scramblase is shown in green, with its associated subunit Basigin in yellow. Upon activation of caspase 3 (3GJS), Xkr8-mediated PS exposure at the cell surface is activated*

*through cleavage of a C-terminal peptide (in red), which is concurrent with inhibition of ATP11C-mediated inward movement of PS through cleavage of its large cytosolic loop (in red) by caspase 3. Xkr4-mediated PS exposure requires both direct cleavage by caspase 3, which removes a large C-terminal domain (in red) and additional activation through binding of a C-terminal fragment of the nuclear protein XRCC4 (in purple), released upon cleavage by caspase 3 (cleavage site in red). The portion of the XRCC4 C-terminal fragment that binds Xkr4 is colored red.*

Exposure of PS on the cell surface may also result from scramblase activation. TMEM16F and several other TMEM16 family members function as $Ca^{2+}$-dependent lipid scramblases and are responsible for exposing PS in activated platelets, whereas Xkr family members Xkr9, 8 and 4 have been found to stimulate PS scrambling under apoptotic conditions [41]. TMEM16 proteins function as homodimers. Each protomer contains a hydrophilic pathway that would shield lipid headgroups, the access to which is controlled by $Ca^{2+}$ binding to a site buried in the hydrophobic core of the membrane (Figure 2). Transport of lipids through the membrane may be facilitated by the ability of certain TMEM16 proteins to distort the membrane bilayer, thereby lowering the energy barrier for lipid flip-flop [42].

Recent studies revealed the architecture of Xkr proteins Xkr8 and Xkr9 [*43,*44]. Whereas Xkr8 interacts with basigin, a type I membrane protein, no evidence for such interaction was shown for Xkr9 (Figure 2). The activity of Xkr proteins is tightly regulated by apoptotic signals. Both Xkr8 and 9 are positively regulated by caspase-mediated cleavage of their C-terminal region. In addition, caspases act as inhibitors of lipid transport by cleaving the P4-ATPase ATP11C (Figure 2). Recently, Maruoka and colleagues uncovered an elaborate regulatory mechanism for Xkr4, whereby a caspase-cleaved fragment of XRCC4, a nuclear protein involved in DNA repair, acts as an activating factor of caspase-cleaved Xkr4 [**45].

Interestingly, a constitutively active Drosophila homolog of Xkr8 promotes high elasticity and low tension of insect cell membranes, indicating that transbilayer lipid dynamics controls membrane deformability. Scramblase-catalyzed lipid flip-flop may also regulate membrane deformation in mammalian cells [*46]. Another recent study suggests that dynamic PS flip-flop may occur during cell migration [47]. A higher cytosolic surface charge was observed in PM protrusions at the front of migrating cells compared to the back, which correlated with a transient increase in PS in the outer PM leaflet; the molecular mechanism of this flip-flop remains to be explored.

**Partnership between lipid scramblases and LTPs**

Apart from TMEM16 and Xkr proteins, other proteins have been shown to facilitate phospholipid scrambling, including the ER-resident proteins TMEM41B and VMP1, which regulate LD and lipoprotein biogenesis, and the ATG9A autophagy protein present in Golgi-derived vesicles [48-50]. Partnership between these scramblases and the channel-like LTP ATG2A, which connects the ER and the autophagosomal membrane and interacts with TMEM41B, VMP1 and ATG9A, has been proposed to be important for generating new membrane [**48]. It is not known what powers the transport of phospholipids towards the expanding membrane, for example during autophagosome formation; an enticing candidate is lipid synthesis at the ER [51].

There is no evidence for PS specificity in these pathways. However, depletion of TMEM41B and VMP1 results in a slight change in PS distribution [50], and a recent study suggests that PS enrichment occurs during maturation of the yeast autophagosomal membrane [52]. The importance of PS in non-canonical autophagy has been recently shown by Durgan and co-workers, who have found that ATG8 can be conjugated to PS instead of PE thanks to a switch in the lipid specificity of the conjugating enzyme [*53].

A functional partnership relevant for PS distribution is illustrated by the interaction between the yeast ER-embedded TMEM16 protein Ist2, which tethers the ER and the PM, and Osh6. Upon deletion of Ist2, cellular PS levels decrease, probably due to inhibition of the PS synthase Cho1 by its own product [54]. Moreover, functional partnership may arise in the context of endosome fission, where both ORP10 and ATP8A1 contribute to PS enrichment in the cytosolic leaflet of endosomes, which facilitates recruitment of the fission protein EHD1 [12].

Similar partnership between scramblases and lipid-transfer proteins has been shown in the context of inflammation, where high extracellular ATP concentration induces PS exposure at the surface of T lymphocytes. The missing link between elevated ATP and PS exposure has now been unraveled: a complex formed by Xk and the LTP VPS13A at ER-PM contact sites was shown to be essential for P2X7-mediated PS exposure [55,56].

**PS nanodomains at the PM**

PS can also laterally segregate in membranes. It is proposed that enrichment of PS at membrane contact sites is important for its export from the ER [7,57], but this has not been directly demonstrated. The distribution of PS at the PM can be more directly assessed, and the importance of dynamic clustering of PS into nanodomains for signaling at the PM is well established. The small GTPase K-RAS, which regulates cell proliferation through the mitogen-activated protein kinase pathway and is mutated in a large number of cancers, binds to the PM via a bi-partite C-terminal signal composed of a lysine patch and a highly branched farnesyl lipid anchor. K-RAS localization depends on the activity of the PS transfer proteins ORP5 and ORP8, and on the PI4-kinase that generates PI4P used in PS exchange [58] (Figure 3). At the PM, K-RAS forms dynamic nano-clusters, which are required for its signaling activity and are dependent on PS. K-RAS recognizes both the headgroup and specific acyl chain composition of PS, preferring mixed-chain PS species containing one saturated and one unsaturated acyl chain [59,60]. The PS acyl chain specificity is dependent on the precise chemistry of the K-RAS lipid anchor. PS can associate with sterols, likewise in an acyl-chain selective manner [61], and sterol distribution at the PM is affected by PS levels [16,62,63]. Glycosphingolipids, which reside in the outer PM leaflet, can also promote PS clustering and K-RAS signaling (Figure 3). This association specifically depends on the presence of long acyl chains in both lipids [64], in agreement with a previous observation that PS species with long acyl chains can promote trans-bilayer coupling of two membrane leaflets [65].

Together, K-RAS signaling is a good illustration of how the combination of small affinities can contribute to a strong preference for PS retention at the PM and promote its lateral nano-clustering (Figure 3). Another interesting example is the Rho-of-plants (ROP) GTPase ROP6, which contains a hyper-variable C-terminal tail similar to K-RAS. ROP6 forms nanoclusters at the PM in a PS-dependent manner, which are critical for hormone signaling and osmo-sensing in plant cells [66]. A recent study shows that the PM receptor kinase FERONIA (FER) also controls ROP6 nano-cluster dynamics and downstream signaling *via* controlling PS distribution and levels at the PM [67].

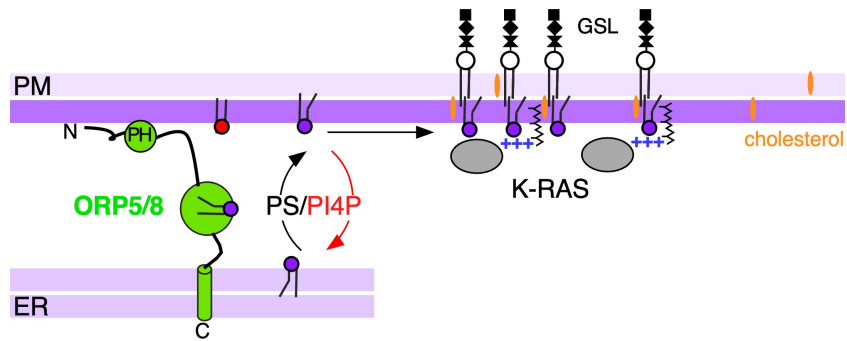

**Figure 3. Different mechanisms promote PS clustering in the cytosolic leaflet of the PM and K-RAS signaling.** *PS (in purple) is delivered to the PM by the LTPs ORP5 and ORP8, which exchange it with PI4P (in red) that is produced locally by PI4-kinase. ORP8 has a preference for unsaturated PS [10]. K-RAS interacts with PS headgroup and acyl chains via its C-terminal localization signal, composed of a lysine patch and a branched lipid anchor. PS and K-RAS nano-clustering is also promoted by interactions with cholesterol, which depend on acyl chain unsaturation, and interactions with glycosphingolipids (GSL) in the outer PM leaflet, which depend on acyl chain length [58,60,64].*

## Conclusion

Recent years have brought an exponential increase in studies of cellular lipid traffic and its implications for other cellular processes, as attested by the large number of very recent papers cited in this review (and others that we could not cite due to space constraints). However, understanding of lipid flow in the cell requires quantitative analysis at high (nanometer range) spatial as well as temporal resolution. This is far from an easy task, given the difficulty in labeling and tracking lipids, their molecular diversity, where seemingly small differences can have a large effect [60,61], and the diverse (often low affinity) molecular interactions that contribute to their behavior. It is in fact striking how much of our knowledge on intracellular distribution of PS and other lipids stems from careful biochemical and cell fractionation studies performed several decades ago, long before the explosion in lipidomics, high-resolution imaging techniques and the use of fluorescent probes in cells [27,57]. The latter techniques have brought many new insights, but also controversy (the continuing debate on the nature and functional importance of lipid rafts in cells being just one obvious example). The understanding of PS function and flow in the cell necessarily requires diverse approaches and revisits of old discoveries that may be perceived as incremental (or obvious), but are absolutely essential, given how fine-tuning of various lipid gradients and interactions, coupled with feedback loops, can lead to vastly different outcomes.

## Acknowledgments

We wish to thank Bruno Antonny and Guillaume Drin for insightful discussions. This work was supported by the Agence Nationale de la Recherche Grant (ANR-20-CE13-0030-02), the CNRS, Université Paris-Saclay, and FRM (ARF202209015714) grant to T.D.

## Conflict of Interest

The authors declare no conflict of interest.